\documentclass[fleqn,usenatbib]{mnras}

\usepackage{newtxtext,newtxmath}
\usepackage{hyperref}
\usepackage[T1]{fontenc}
\usepackage{lscape}
\DeclareRobustCommand{\VAN}[3]{#2}
\let\VANthebibliography\thebibliography
\def\thebibliography{\DeclareRobustCommand{\VAN}[3]{##3}\VANthebibliography}

\usepackage{graphicx}	% Including figure files
\usepackage{amsmath}	% Advanced maths commands
\title[O($^3P$) + N$_2$ scattering cross-sections]{Quantum scattering cross-sections for O($^3P$) + N$_2$ collisions for planetary aeronomy}

\author[Kumar, Kumar, and Yamijala]{
Sanchit Kumar,$^{1,2,}$\thanks{Email: \url{sanchitchhabra37@gmail.com} (Sanchit K.) \hfill \break
\hspace*{1.1cm}\url{yamijala@iitm.ac.in} (S.S.R.K.C.Y)}
Sumit Kumar,$^{1,2}$
Marko Gacesa,$^{3}$ 
Nayla El-Kork,$^{3,4}$
and Sharma S. R. K. C. Yamijala$^{1,2,5,6, \textcolor{blue}{\star}}$
\\
$^{1}$Department of Chemistry, Indian Institute of Technology Madras, Chennai 600036, India \\
$^{2}$Centre for Quantum Information, Communication, and Computing, Indian Institute of Technology Madras, Chennai 600036, India \\
$^{3}$Department of Physics, Khalifa University, P.O. Box 127788, Abu Dhabi, United Arab Emirates. \\
$^{4}$Space and Planetary Science Center, Khalifa University, Abu Dhabi, United Arab Emirates. \\
$^{5}$Centre for Atomistic Modelling and Materials Design, Indian Institute of Technology Madras, Chennai 600036, India \\
$^{6}$Centre for Molecular Materials and Functions, Indian Institute of Technology Madras, Chennai 600036, India \\
}

\date{Accepted XXX. Received YYY; in original form ZZZ}

\pubyear{2023}

\begin{document}
\label{firstpage}
\pagerange{\pageref{firstpage}--\pageref{lastpage}}
\maketitle

\begin{abstract}
“Hot atoms”, atoms in their excited states, transfer their energy to the surrounding atmosphere through collisions. This process (known as thermalization) plays a crucial role in various astrophysical and atmospheric processes. Thermalization of hot atoms is mainly governed by the amount of species present in the surrounding atmosphere and the collision cross-sections between the hot atoms and surrounding species. In this work, we investigated the elastic and inelastic collisions between hot oxygen atoms and neutral N$_2$ molecules, relevant to oxygen gas escape from the Martian atmosphere and for characterizing the chemical reactions in hypersonic flows. We conducted a series of quantum scattering calculations between various isotopes of O($^3P$) atoms and N$_2$ molecules across a range of collision energies (0.3 to 4 eV), and computed both their differential and collision cross-sections using quantum time$-$independent coupled-channel approach. Our differential cross-section results indicate a strong preference for forward scattering over sideways or backward scattering, and this anisotropy in scattering is further pronounced at higher collision energies. By comparing the cross-sections of three oxygen isotopes, we find that the heavier isotopes consistently have larger collision cross-sections than the lighter isotopes. As a whole, the present study contributes to a better understanding of the energy distribution and thermalization processes of hot atoms within atmospheric environments. Specifically, the cross$-$sectional data presented in this work is directly useful in improving the accuracy of energy relaxation modeling of O and N$_2$ collisions over the Mars and Venus atmospheres.

\end{abstract}


%%%%%%%%%%%%%%%%%%%%%%%%%%%%%%%%%%%%%%%%%%%%%%%%%%

%%%%%%%%%%%%%%%%% BODY OF PAPER %%%%%%%%%%%%%%%%%%

\section{Introduction}
%Why studying gases is important? How to accurately model them?
The study of the kinetic behavior of gases is crucial for simulating and understanding various gaseous environments, including those found in Earth's and other planetary atmospheres, combustion reactions, plasma chemistry, and high-speed aerodynamics, among others \citep{pintassilgo2016power, pavlov2011vibrationally, doi:10.1021/acs.jpca.9b05675, yankovsky2020model, keidar2015plasma}. To model the kinetic behavior of gases accurately, it is necessary to study the elastic, inelastic, and dissociation processes that occur due to the collisions among the gaseous species. \citep{https://doi.org/10.1029/2000JA000085, doi:10.1021/cr400145a,doi:10.1063/1.5058126} Among the different kinds of collisions, studying the collisions between atomic and diatomic species is crucial as it provides insights into important processes like thermalization of "hot atoms" and escape rates of gaseous species from the upper atmospheres of planets \citep{GACESA201790}.

%Non-equilibrium conditions often occur in molecular systems, requiring the use of detailed state-to-state models to understand and control technology-relevant phenomena. Input data for these models can be sourced from both experimental results and theoretical calculations.
%Experimentally, the molecular beam technique is widely recognized as a highly effective means of gathering information about the potential energy between two particles through the measurement of collision cross-sections. Other methods, such as spectroscopy and the study of properties in gaseous, liquid, and solid states, can also provide information about the potential energy. However, each of these methods has its own limitations and can only provide information about specific aspects of the potential energy function, rather than the entire range of potential energy.

%Why studying energetic atoms is important? Where do they exist? How do they lose their energy?
Hot atoms are formed in the upper atmospheres of planets through several mechanisms like dissociative recombination, collisional quenching, photo-dissociation, and photo-electron impact dissociation. In general, the excess energy of these atoms is dispersed to the surrounding atmospheric gases through both elastic and inelastic (can be reactive or non-reactive) collisions. In a non-reactive inelastic collision, hot atoms excite the colliding molecules to higher rovibrational levels, and in a reactive inelastic collision, new chemical reactions (involving the making or breaking of bonds) are triggered.

%Which energetic atoms were studied earlier? Why did you consider Hot O atoms for your study, and why specifically on Mars?
Several earlier works have mainly focused on how hot N($^4$S) atoms are generated and how they attain thermal equilibrium in planetary atmospheres \citep{solomon1983possible, kharchenko1997kinetics, lie1991kinetics, gerard1991non, sharma1996production}. However, apart from hot N atoms, various planetary atmospheres also contain energetic oxygen atoms \citep{https://doi.org/10.1002/2015JE004890}. For example, in the upper thermosphere and exosphere of Earth, evidence has shown that a hot corona composed of oxygen exists \citep{oliver1997hot, bisikalo1995kinetic, hickey1995new, yee1980detection}. Similarly, energetic O atoms are present in Mars' thermosphere \citep{https://doi.org/10.1029/2009JE003388}. Indeed, the generation of hot O atoms is one of the ways through which O$_2$ gas escapes from Mars. Since understanding the escape of oxygen provides insights into how the atmosphere of Mars has changed over time, it is beneficial to study the collision reactions between hot O atoms and other molecules present in the Mars atmosphere.

%How hot oxygen atoms are generated on Mars? What are the molecules with which the hot O atoms would be colliding with?
Based on the current understanding, the source of hot O atoms on Mars is primarily due to the dissociative recombination of O$_2^+$ ions  \citep{https://doi.org/10.1029/90JA01676, https://doi.org/10.1029/2007JE002915, lammer2013outgassing, https://doi.org/10.1029/2009JE003388, YAGI2012682}. The mechanism is non-thermal and it generates hot O atoms of varying energies through one of the following pathways:
%it generates hot O atoms of varying energies (above the energy required for their escape)
\begin{align*}
                  {O_2}^{+} + e^{-}  \rightarrow  O(^3P) +  O(^3P) + 6.96  eV \\
                                     \rightarrow  O(^3P) +  O(^1D) + 5.00  eV \\
                                     \rightarrow  O(^1D) +  O(^1D) + 3.02  eV \\
                                     \rightarrow  O(^1D) +  O(^1S) + 0.80  eV \\
\end{align*}
The initial two processes in the reaction generate highly energetic particles, referred to as superthermals, that can escape the gravitational pull of Mars and enter into space. However, their successful escape is contingent on them avoiding collisions with other molecules in the atmosphere, which can lead to a partial loss of their energy and momentum. For example, in the thermosphere of Mars, hot O atoms experience frequent collisions with either other hot O atoms or CO$_2$ molecules, which are the most common collision partners in that region \citep{https://doi.org/10.1029/2009JE003388}. Other species that also have significant contributions in determining the energy distribution within the hot O corona are CO and N$_2$. Indeed, \cite{https://doi.org/10.1002/2015JE004890} found that adding CO and N$_2$ to the thermosphere (along with O and CO$_2$) reduces the escape rate of O atoms by 40\%. Recently, \cite{chhabra2023quantum} computed cross-sections for the O($^3P$)-CO complex. However, in determining the total escape rate of hot O atoms from Mars, O($^3P$)-N$_2$ collision cross-sections also play a significant role \citep{https://doi.org/10.1002/2015JE004890}.

% What are the uses/applications of studying N2 molecules' collision with O? %What are some of the earlier simulations on O + N2 collisions?
Apart from Mars, the O-N$_2$ cross-sections also play a crucial role in the oxygen-rich environment models of Venus \citep{https://doi.org/10.1029/2010JE003697}. Also, these collisions are of particular interest to environmental chemistry and hypersonic flight regimes. Considering this importance, O($^3$P) + N$_2$ collisions have been studied experimentally and theoretically by several authors \citep{hong2021reconciling, yokota2021effect, https://doi.org/10.1029/98JA02198, https://doi.org/10.1029/2002JA009566, upschulte1992infrared, https://doi.org/10.1029/94JA01745, walch1987calculated, gilibert1992dynamics,bose1996thermal, koner2020accurate, doi:10.1021/acs.jpca.7b04442, doi:10.1021/acs.jpca.1c10346}. \cite{walch1987calculated} conducted CASSCF/CCI calculations using extensive Gaussian basis sets to examine certain segments of the potential energy surfaces associated with the Zeldovich mechanism, which involves the transformation of N$_2$ into NO. Similarly, \cite{doi:10.1021/acs.jpca.7b04442} conducted comprehensive simulations of atomic oxygen interacting with molecular nitrogen in their respective fundamental electronic states and reported the reactive, inelastic, and dissociative processes. Using molecular dynamics simulations, \cite{hong2021reconciling} simulated the inelastic collisions between atomic oxygen and molecular nitrogen, and obtained a remarkable level of quantitative agreement with experimental results of total relaxation rates. \cite{doi:10.1021/acs.jpca.1c10346} investigated the energy transfer and dissociation processes of an N$_2$ + O system by employing rovibrational-specific quasi-classical trajectory (QCT) and master equation analyses. 

Although N$_2$ collision with O($^3$P) has been studied by many groups, only \cite{https://doi.org/10.1029/98JA02198} has computed the elastic and total integral cross-sections over a range of energies. It is important to note that, the knowledge of both elastic and inelastic cross-sections is essential in determining the escape rate of a gas from the planetary atmosphere \citep{https://doi.org/10.1002/2016JA023525}. Moreover, to obtain a precise depiction of the atmosphere, it is crucial to consider not only the overall cross-section but also the angular and energy dependence of the cross-section \citep{10.1063/1.1734304,10.1063/1.452944}, which were not considered in \cite{https://doi.org/10.1029/98JA02198} work (specifically, elastic differential cross-sections are absent). As such, currently, there is no existing differential cross-section data for elastic collisions involving O($^3$P) and N$_2$. Also, the study  performed by \cite{https://doi.org/10.1029/98JA02198} used the potential energy surface (PES) of \cite{gilibert1992dynamics}, which was generated using the CASSCF/CCI method, a low-level theory as per the current standards. Moreover, these calculations were only performed on specific sections of the PES rather than on the entire PES. Considering the above issues with earlier works, in this article, we investigated the quantum-mechanical scattering cross-sections of $^{16}$O($^3$P) + N$_2$, $^{17}$O($^3$P) + N$_2$, and $^{18}$O($^3$P) + N$_2$ systems, considering both elastic and inelastic interactions, over a range of collision energies from 0.3 to 4 eV using accurate PESs generated using MRCI+Q method (see methodology section for further details). Furthermore, we also calculated the differential and total integral cross-sections that are significant for the transportation and thermalization of hot O atoms in non-thermal equilibrium conditions (present in the upper and middle atmospheres of planets).

%\begin{figure}  %[!ht] 
%\begin{center}
%\includegraphics [height=0.30\textwidth]{n2o.eps}
%{\caption{\label{fig:coordinate}Coordinate system for O($^3$P)+ N$_2$ complex.}}
%\end{center}
%\end{figure}
%---------------------------------------------------------------------------------------

\section{Methodology}
\subsection{Potential Energy Surfaces (PESs)}
To compute the quantum-mechanical scattering cross-sections of "O($^3$P) + N$_2$", we need accurate PESs of O + N$_2$ system, which mainly contain the triplet states of N$_2$O, namely, 3A$'$ and 3A$''$. In the literature, there are multiple PESs of N$_2$O, corresponding to the calculations at various levels of theory \citep{walch1987calculated, gamallo2003ab, lin2016global, denis2017reactive, gamallo2003quantum}. Recently, \cite{koner2020accurate} generated accurate PESs of N$_2$O's 3A$'$ and 3A$''$ states using the multi-reference configuration interaction (MRCI) method along with the Davidson correction (Q), where the correction ensures the size consistency that is absent in MRCI. Also, they used augmented Dunning-type correlation consistent polarized triple zeta (aug-cc-pVTZ) basis sets in their calculations. They validated the accuracy of their PESs by computing the reaction rates for N$_2$ + O and NO + N collisions (using QCT simulations), which are in excellent agreement with the experimentally obtained rates. Considering these results, we used the 3A$''$ PES of  \cite{koner2020accurate} in our calculations. We omitted the 3A$'$ PES, since we find its contribution to the cross-section is negligible (for all the collision energies considered in this work). Moreover, earlier work by \cite{https://doi.org/10.1029/98JA02198} also found that 3A$'$ PES has no significant contribution to the cross-section.

\subsection{Collisional Dynamics}
Next, to compute the collision cross-sections of N$_2$-O using the 3A$''$ PES, we used the quantum close-coupling (CC) equations, developed by \cite{doi:10.1098/rspa.1960.0125}. In their derivation, the  vibrational degrees of freedom are ignored, and  the  diatomic  molecule (here, N$_2$) is  treated as  a linear rigid  rotor. The expression for the rotational cross-section ($\sigma_{j\rightarrow j'}$) associated with a transition from an initial rotational state, $j$, to a final rotational state, $j'$, involves the utilization of scattering matrix elements S$_{jj'll'}$ as,

%The aim of this paper is the use of the fitted N$_2$-O PES to obtain the cross-sections. The quantum close-coupling (CC) equations are introduced into the field of molecular collisions by \cite{doi:10.1098/rspa.1960.0125}. The rotational cross-sections for transitions from an initial rotational state denoted by $j$ to a final rotational state denoted by $j'$ can be expressed in the form of the scattering matrix S$_{jj'll'}$,
\begin{multline}{\label{eq01}}
\sigma_{j\rightarrow j'}(E_k) = \frac{\pi}{k_{j}^2(2j+1)}\sum_{J=0}^{J_{max}}(2J+1) \\
 \times \sum_{l=|J-j|}^{J+j} \sum_{l'=|J-j'|}^{J+j'}
 |\delta_{jj'} \delta_{ll'} - S_{jj'll'}^{J}(E_k)|^2
\end{multline}

where $k_{j}$ = $\sqrt{2\mu E_{k}}/{\hbar}$ denotes the wave vector of the incoming channel, and $E_k$ is the kinetic energy of the incident particle (here, O), and it can be expressed as, $E_k$ = $E$ $-$ $E_j$, where $E$ is the total energy of the N$_2$ + O system, and $E_j$ the rotational energy of the N$_2$ molecule. $l$ and $l'$ are the initial and final orbital angular momenta of the collision complex, respectively. Finally, $\textbf{J}$ represents the combined angular momentum, expressed as $\textbf{J}$ = $\textbf{l}$ + $\textbf{j}$, where $\textbf{l}$ represents the orbital angular momentum of the collision complex, and $\textbf{j}$ represents the rotational angular momentum of the N$_2$ system. Equation (1) is widely used in practical applications, where the sum over $J$ is limited to $J_{max}$. The value of $J_{max}$ is chosen carefully by considering the contribution from terms with higher J values to the cross-section, and is truncated at a point where their effect is no longer significant.

There are several approximate methods that can be employed to address concerns related to computational time and accuracy. The computational time is reduced in the coupled state (CS) approximation, as suggested by \cite{doi:10.1063/1.1681388}, through the omission of Coriolis coupling between different $\Omega$ values. Here, $\Omega$ represents the projection of the diatom's angular momentum quantum number along the body-fixed axis. We have used CS approximation to compute the cross-sections.

The integral cross-section, within the CS formalism, is given by

\begin{multline}{\label{eq02}}
\sigma_{j\rightarrow j'}(E_k) = \frac{\pi}{k_{j}^2(2j+1)} \sum_{J=0}^{J_{max}}(2J+1)\\
 \times \sum_{\Omega=0}^{\Omega_{max}}(2-\delta_{\Omega 0})
 |\delta_{jj'} - S_{jj'}^{J \Omega}(E_k)|^2
\end{multline}

where $\Omega_{max}$ = 0,  1 , 2 , 3, ....,max($J,j$).

The total cross-section for the complete scattering process starting from a specified initial state is expressed as:

\begin{equation}{\label{eq03}}
\sigma_{j}(E_k) = \sum_{j'}\sigma_{j,j'}(E_k)
\end{equation}

where $j'$ = 0,....$j_{max}$. 

The differential cross-sections (DCSs) are determined by:

\begin{equation}{\label{eq04}}
Q_{jj'}(\theta,E) = d\sigma_{j,j'}(\theta,E)/d\Omega
\end{equation}

The total DCSs for a particular initial state are obtained by integrating the individual DCSs over all possible final scattering angles.
 \begin{equation}{\label{eq05}}
Q_{j}(\theta,E) = \sum_{j'}d\sigma_{j,j'}(\theta,E)/d\Omega
\end{equation}

In this context, d$\Omega$ = sin $\theta$ d$\theta$ d$\phi$ represents the solid angle element, where $\theta$ is the angle of scattering between the O atom's center-of-mass velocity vector and the N$_2$ molecule. Notably, $\theta$ = 0$^{\circ}$ and 180$^{\circ}$ are associated with forward and backward scattering, respectively.
The rotational energies of N$_2$ molecule are calculated using a value of 1.998241 cm$^{-1}$ for the rotational constant \citep{doi:10.1063/1.2436891}. Due to its symmetry, N$_2$ molecule obeys ${\Delta j}= \pm 2$ rules.  

\subsection{Calculation Details}

\begin{table} %[h]
\caption{\label{tab:table1}Inputs utilized in the MOLSCAT calculations.}
\begin{tabular}{cc}
\cline{1-2}
$\mu_{16}$ = 10.183 a.m.u     & $\mu_{17}$ = 10.579 a.m.u    \\
$\mu_{18}$ = 10.958 a.m.u     & $R_{min}$ = 0.8 {\AA}        \\
$R_{max}$ = 20 {\AA}          & RVFAC = 1.3                  \\
DTOL = 0.001 {\AA$^2$}        & OTOL = 0.005{\AA$^2$}        \\ \cline{1-2}     
\end{tabular}
\end{table}

The MOLSCAT code \citep{hutson1994molscat,HUTSON20199} was utilized to determine state-to-state quantum mechanical cross-sections (both elastic and inelastic) for the scattering process involving O($^3$P) + N$_2$ ($j$) $\rightarrow$ O($^3$P) + N$_2$ ($j'$). The rotational quantum numbers, $j$ and $j'$, represent the initial and final quantum states of the N$_2$ molecule, respectively. We solved a set of time-independent Schrodinger equations that were coupled-channel in nature, using a close-coupling formalism. The modified log-derivative Airy propagator was employed to integrate the coupled-channel equations \citep{doi:10.1063/1.452154}. The N$_2$ molecule was modeled as a rigid rotor, and the interaction with O($^3$P) was represented by the potential energy surfaces of \cite{koner2020accurate}. The N$_2$ bond distance was fixed at an equilibrium distance of 2.0743 a$_0$ \citep{herzberg2013molecular}. In the angular expansion of the potential, we included 22 Legendre terms, where the expansion coefficients were determined using a 23-point Gaussian quadrature, using the standard VRTP mechanism of the MOLSCAT. The scattering problem was solved at different collision energy values that were ranging from 0.3 to 4 eV. Reduced mass of N$_2$-O is 10.183 a.m.u. for $^{16}$O-N$_2$, 10.579 a.m.u. for $^{17}$O-N$_2$, and 10.958 a.m.u. for $^{18}$O-N$_2$. A comprehensive series of tests were performed to guarantee convergence with respect to both the size of the basis set and the numerical integration parameters. To ensure that elastic cross-sections for the N$_2$ molecule reach satisfactory convergence, we utilized a basis set consisting of rotational functions, where the number of basis functions included is defined by $j_{max}$. In general, for this system, we observed that elastic cross-sections converge for $j_{max}$ = 30. Moreover, we observed a similar level of convergence even when employing the centrifugal sudden (CS) approximation by keeping JZCSMX = 2 (which simplifies the calculations by neglecting transitions involving rotational states higher than j = 2). On the other hand, to ensure the convergence of inelastic cross-sections for initial rotational states j (which are present in the Maxwell–Boltzmann tail in the thermal upper atmosphere of Mars), we conducted production runs using a significantly larger basis set with $j_{max}$ = 70 and JZCSMX = 8. For the larger rotational basis set, the calculation was not feasible without invoking the CS approximation. To guarantee convergence of the cross-sections, we incorporated an ample number of the total angular momentum partial waves $J$. The value of $J_{max}$ was observed to increase with the collision energy. The maximum achieved value of $J$ was 1816. All the important parameters that we used for the MOLSCAT calculations are given in Table \ref{tab:table1}.

%This figure belongs to results and discussion
\begin{figure}  %[!ht] 
\begin{center}
\includegraphics [height=0.35\textwidth]{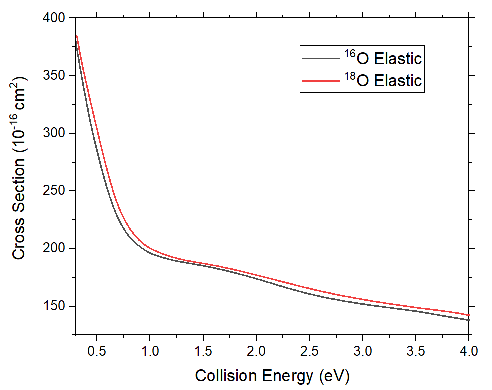}
{\caption{\label{fig:elastic_16_18}Elastic cross-sections for $^{16}$O+ N$_2$ and $^{18}$O+ N$_2$ varying with collision energy.}}
\end{center}
\end{figure}

\section{Results and Analysis}
First, we presented the elastic cross-section data for the $^{16}$O+ N$_2$ and $^{18}$O+ N$_2$ collisions, in accordance with collision energy, in Figure \ref{fig:elastic_16_18}. Clearly, the cross-section increases with a decrease in the collision energy, which is a general feature \citep{10.1093/mnras/staa1086,D1CP04273D}. Moreover, the cross-sections of both these isotopes have a similar energy trend. However, interestingly, the cross-section of $^{18}$O $-$N$_2$ is slightly larger than that of the $^{16}$O $-$N$_2$. 
%For a given kinetic energy, the larger cross-section of $^{18}$O $-$N$_2$ could be mainly due to its heavier reduced mass.
We have also computed the cross-section data of $^{17}$O-N$_2$ and it lies between the $^{16}$O+ N$_2$ and $^{18}$O+ N$_2$ results. Together, we can conclude that the cross-sections for heavier isotopes are slightly larger than those of the lighter isotopes. %due to their heavier reduced masses. 
Our results are in good agreement with the earlier studies. For example, using numerical scattering studies, \cite{https://doi.org/10.1029/2010GL045763} studied the atom-atom collisions between $^{3}$He/$^{4}$He and O, and obtained larger cross-sections for the $^{4}$He-O scattering than for the $^{3}$He-O (in the center of mass frame). Similarly, using both semi-classical and numerical studies, \cite{lewkow2012energy} obtained larger cross-sections for heavier isotopes. However, it should be noted that the relationship between cross-sections and reduced mass is not always straightforward, and it depends on the specific systems. In our case, we observed that the cross-sections for heavier isotopes are slightly larger than those of the lighter isotopes.

By comparing our results with those of \cite{https://doi.org/10.1029/98JA02198}, we find that the elastic cross-sections obtained by us are lower than their cross-sections, which could be attributed to the better PES and the computational method that we used in our calculations.  

%   For energies less than 80 cm$^{-1}$, several resonances can be noticed as shown in figure \ref{fig:low_elastic}. This behavior relates to the decay of quasi bound states constituted by attractive part of the surface. These resonances vanish at higher energies. A similar behavior is also noticed in recent works (cite). \\

\begin{table} %[h]
\caption{\label{tab:table3}Elastic cross-sections ($\nu$=0, $j$=0) (units of 10$^{-16}$ cm$^2$) in accordance with selected energies (in eV) for $^{16}$O $-$N$_2$, $^{17}$O $-$N$_2$ and $^{18}$O $-$N$_2$}.
\centering
\begin{tabular}{cccc}
\cline{1-4}\\
E (eV)  & $^{16}$O $-$N$_2$ & $^{17}$O $-$N$_2$ & $^{18}$O $-$N$_2$ \\ \cline{1-4} \\
0.75    &    216.191    &    220.982  & 225.686    \\
1.25    &    189.051      &    189.957  & 190.858  \\
1.50    &    185.358      &    186.651   & 187.354  \\
1.75    &    180.294       &    181.671  & 182.976  \\
2.25    &    166.868       &    169.043  & 171.297   \\
2.75   &    155.650      &    157.351  & 159.262   \\
3.5 &    145.978       &    147.688 &  148.612  \\
3.75 &    141.722     &      144.250  & 146.337  \\
4.0 &    137.975     &      140.102  &  142.353 \\
\cline{1-4}          
\end{tabular}
\end{table}

%\begin{table} %[h]
%\caption{\label{tab:table3}Elastic cross-sections ($\nu$=0, $j$=0) (units of 10$^{-16}$ cm$^2$) in accordance with selected energies (in eV) for $^{16}$O with N$_2$}.
%\centering
%\begin{tabular}{cccc}
%\cline{1-4}\\
%E (eV)  & cross-sections & E (eV) & cross-sections \\ \cline{1-4} \\
%0.75    &    216.191     &  2.5   &    160.406     \\
%1.00    &    196.069     &  2.75  &    155.650     \\
%1.25    &    189.051     &  3.00  &    151.863     \\
%1.50    &    185.358     &  3.25  &    148.751     \\
%1.75    &    180.294     &  3.50  &    145.978     \\
%2.00    &    174.019     &  3.75  &    141.723     \\
%2.25    &    166.868     &  4.00  &    137.976     \\
%\cline{1-4}          
%\end{tabular}
%\end{table}

Next, we presented the elastic cross-sections of $^{16}$O $-$N$_2$, $^{17}$O $-$N$_2$ and $^{18}$O $-$N$_2$ varied by selected energies in Table \ref{tab:table3}. To enable a comparison with the data utilized in recent planetary aeronomy studies \citep{fox1993production, LO2021114371,https://doi.org/10.1002/2016JA023461}, we fitted the elastic cross-sections as a function of collision energy ($E_k$) using the functional form $\sigma_n(E_k)=A_n E_k^{b_n}$, where $A_n$ and $b_n$ denote fitting parameters, and $E_k$ is given in eV. For the collision between $^{16}$O and N$_2$, we obtained the parameters $A_n$ and $b_n$ as 2.32 $\times$ 10$^{-14}$ and $-$0.38, respectively. For comparison, we also reported the $A_n$ and $b_n$ parameters for collisions between $^{16}$O and various target species (we took the results from \cite{FOX2018411}) as a function of energy in Table \ref{tab:table4}. This inelastic cross-section data, together with the elastic cross-section data presented in Table \ref{tab:table3} are crucial in calculating the escape rate of O atom from planetary atmospheres.

%\begin{equation}{\label{eq06}}
%\sigma_n(E_k)=A_n E_k(eV)^{b_n}
%\end{equation}

\begin{table} %[h]
\caption{\label{tab:table4}Parameters for elastic cross-sections varying with energy (in eV) for $^{16}$O, with various target species and comparison made by \citep{FOX2018411} using data from below references.}
\centering
\begin{tabular}{cccc}
\cline{1-4}
    Species     &   A                           &  b        & Reference   \\
$^{16}$O-N$_2$  &   2.32  $\times$ 10$^{-14}$   & -0.38     & This work\\
$^{16}$O-Ar     &   1.52  $\times$ 10$^{-14}$   & -0.25     & \cite{doi:10.1063/1.1637343}\\ 
$^{16}$O-CO     &   2.427 $\times$ 10$^{-14}$   & -0.1719   & \cite{chhabra2023quantum}\\ 
$^{16}$O-N$_2$  &   1.98  $\times$ 10$^{-14}$   & -0.0816   & \cite{https://doi.org/10.1029/98JA02198}\\  
$^{16}$O-H$_2$  &   3.084 $\times$ 10$^{-15}$   & -0.2878   & \cite{https://doi.org/10.1029/2012GL050904}\\ 
\cline{1-4}          
\end{tabular}
\end{table}

\begin{figure}  %[!ht] 
\begin{center}
\includegraphics [height=0.40\textwidth]{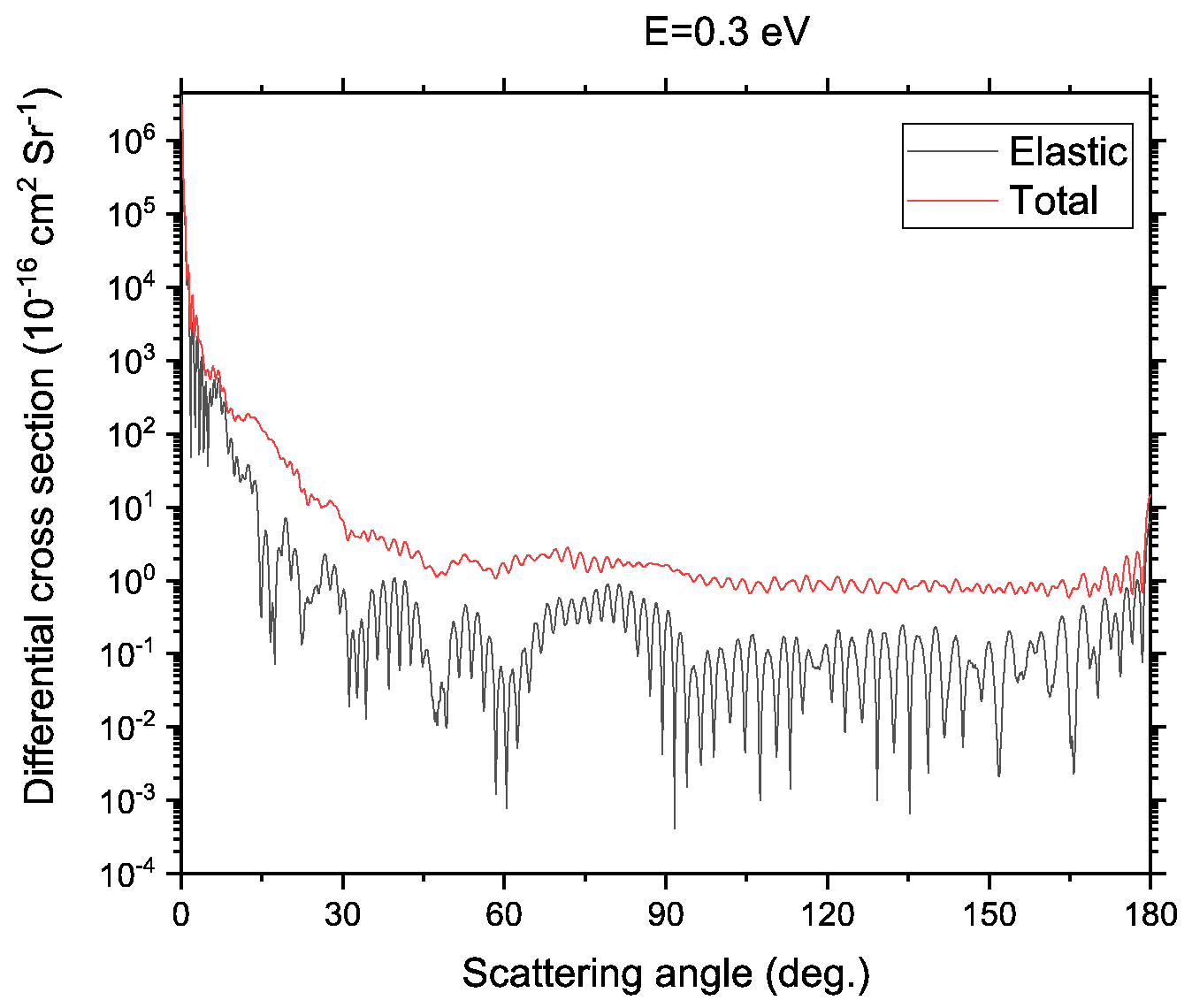}
%\subfloat{\includegraphics [height=0.40\textwidth]{diff_3_5.eps}}
{\caption{\label{fig:elastic_total_03}The dependence of elastic and total differential cross-sections for $^{16}$O+ N$_2$ on the scattering angles at an energy of 0.3 eV.}}
\end{center}
\end{figure}

\begin{figure}  %[!ht] 
\begin{center}
\includegraphics [height=0.40\textwidth]{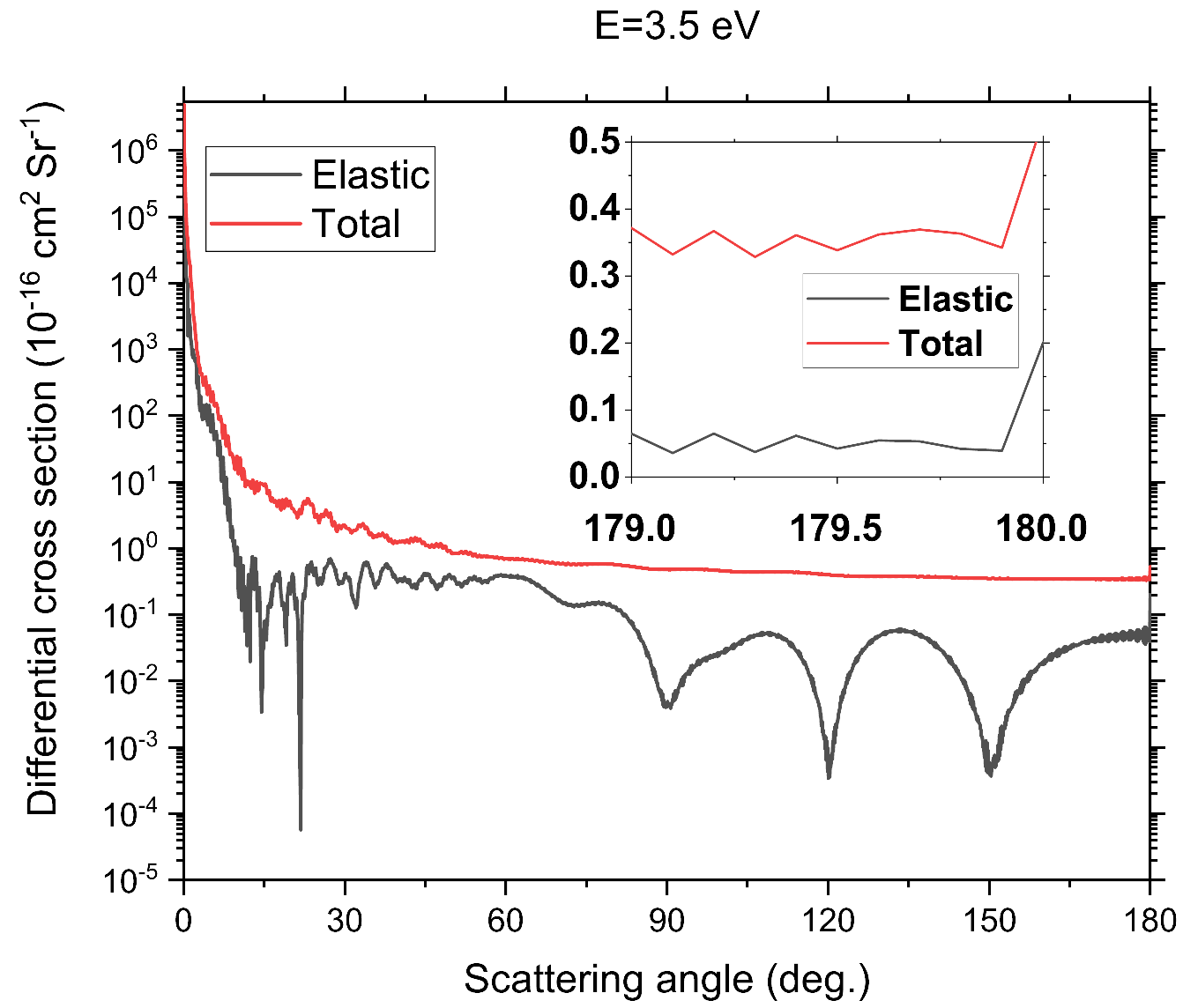}
%\subfloat{\includegraphics [height=0.40\textwidth]{diff_3_5.eps}}
{\caption{\label{fig:elastic_total_35}The dependence of elastic and total differential cross-sections for $^{16}$O+ N$_2$ on the scattering angles at an energy of 3.5 eV.}}
\end{center}
\end{figure}

%--------------------------------------------------------------------------------------
%\subsection{Differential cross-sections}
As mentioned earlier, we performed the scattering calculation for a range of collision energies, and we saved the scattering(S)-matrices for all these collision energies. State-to-state differential cross-sections (DCSs) were computed from these S-matrices. Note that we have calculated DCSs for all the angles from 0$^\circ$ to 180$^\circ$ with step size of 1$^\circ$. In Figures \ref{fig:elastic_total_03} and \ref{fig:elastic_total_35}, we presented the elastic and total differential cross-sections of the $^{16}$O + N$_2$ collision as a function of scattering angle at 0.3 eV and 3.5 eV collision energies, respectively. Next, in Table \ref{tab:table_cross}, we provided the state-to-state cross-sections ($\sigma_{j=0,j'}$) for the $^{16}$O + N$_2$ collision with $j=0$ and $j'= 0$ to $30$ rotational levels.

\begin{figure*}  %[!ht] 
\begin{center}
%\subfloat{\includegraphics [height=0.40\textwidth]{16o_18_o_diff.eps}}
%\subfloat{\includegraphics [height=0.40\textwidth]{16o_18_o_total.eps}}
{\includegraphics [height=0.40\textwidth]{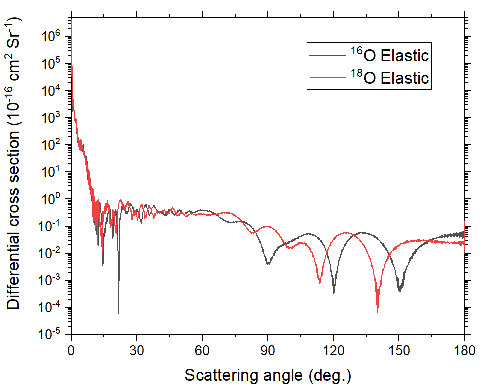}}
{\includegraphics [height=0.40\textwidth]{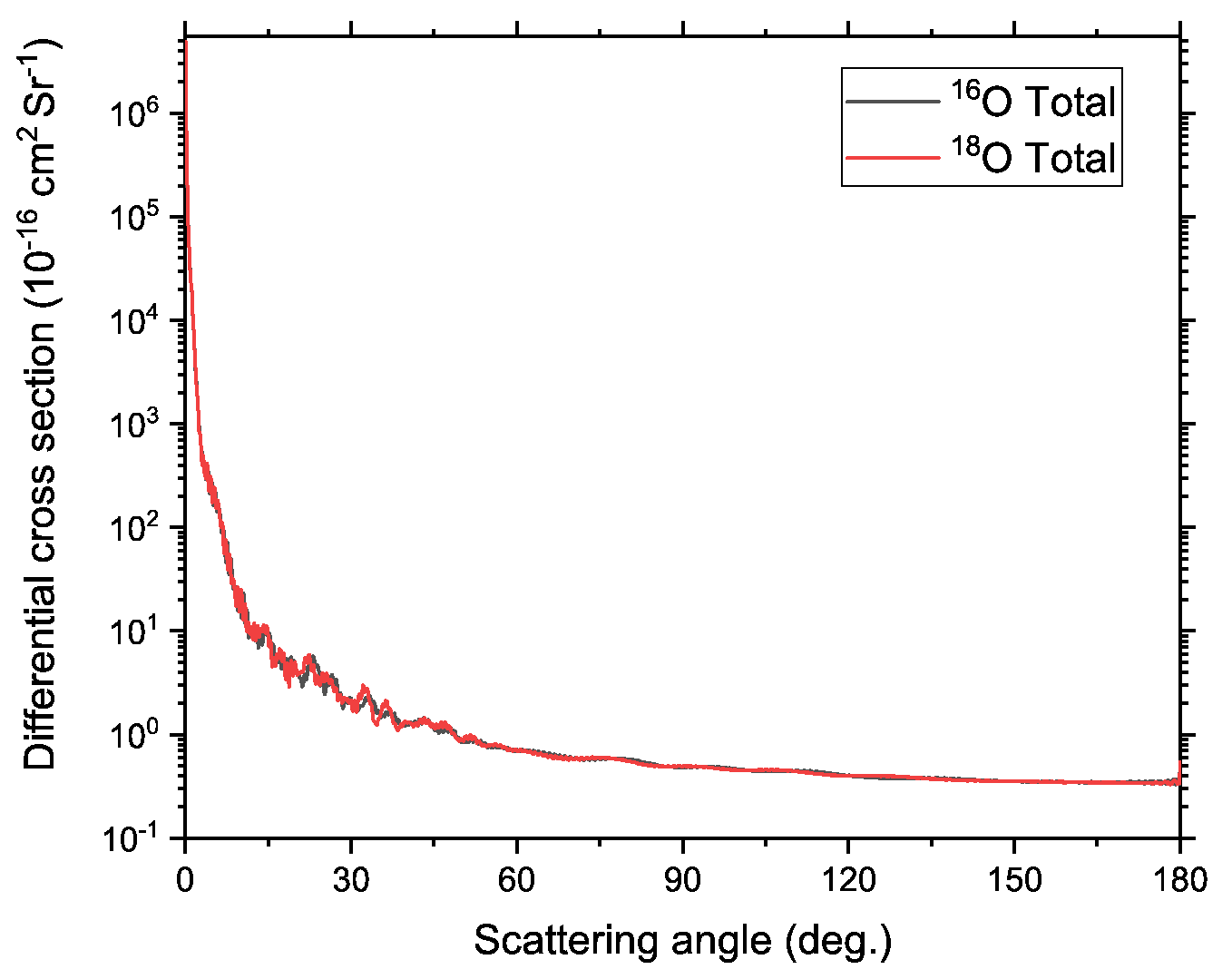}}
{\caption{\label{fig:elastic_total_DCS_16_18}The dependence of elastic and total differential cross-sections for $^{16}$O+ N$_2$ and $^{18}$O+ N$_2$ on the scattering angles at an energy of 3.5 eV.}}
\end{center}
\end{figure*}

Total DCSs (the sum of elastic and inelastic contributions) are strongly anisotropic, having a dominant forward scattering peak (from 0$^\circ$ to 10$^\circ$). Also, as shown in the inset of Figure \ref{fig:elastic_total_35}, one can see a backward scattering peak for large scattering angles. At small scattering angles (up to about $\theta$=10$^\circ$), the DCSs are found to be purely elastic. However, for larger scattering angles, the total DCSs exhibit a magnitude that is up to two orders of magnitude higher than the elastic DCSs, demonstrating that the rotational excitations play a major role at high collision energies. The computed DCSs exhibit rapid oscillations that arise from quantum-mechanical interferences. The presence of the oscillatory structure in elastic cross-sections, especially at low collision energies, is a genuine characteristic that cannot be averaged out. This is because the number of participating partial waves is significantly lower compared to total differential cross-sections that encompass both elastic and inelastic channels. At higher collision energies, the oscillatory structure is eliminated because there are more contributing partial waves involved, which yields smoother cross-sections. The variation in the elastic and total DCSs as a function of scattering angle for $^{16}$O+ N$_2$ and $^{18}$O+ N$_2$ collisions at 3.5 eV are shown in Figure \ref{fig:elastic_total_DCS_16_18}.

%\begin{landscape}
\begin{table*} %[h]
\caption{\label{tab:table_cross}State to state integral cross-sections, ($\sigma_{j=0,j'}$ in the units of 10$^{-16}$ cm$^2$) at selected energies (in eV) for the $^{16}$O collision with N$_2$. $j'=$ is varied from 0 to 30.}
\centering
\begin{tabular}{ccccccc}
\cline{1-7}
$j'$  &    0.750  &     1.500   &   2.25    &   3.25  &   3.75  & 4    \\
\cline{1-7}\\
 0    &  216.192  &   185.358   & 166.869   & 148.751 & 141.723 & 137.975 \\
 2	 &   54.127  &    42.858   &  36.475   & 30.489  & 27.978  & 26.917  \\
 4	 &   19.658  &    46.892   &  56.714   & 56.451  & 55.763  & 54.714  \\
 6	 &   26.089  &    13.929   &  11.373   & 17.549  & 20.054  & 21.243  \\
 8	 &    8.124  &    6.833    &   5.875   &  4.652  &  4.146  &  4.207  \\
10	  &    8.478  &    6.004    &   6.452   &  4.981  &  4.415  &  3.868  \\
12	  &    4.378  &    2.859    &   2.551   &  2.765  &  2.696  &  2.522  \\
14	  &    2.101  &    2.406    &   2.289   &  1.947  &  1.818  &  1.849  \\
16	  &    1.696  &    1.561    &   1.547   &  1.394  &  1.289  &  1.273  \\
18	  &    1.579  &    1.402    &   1.117   &  1.010  &  1.124  &  1.154  \\
20	  &    1.459  &    1.199    &   0.949   &  0.944  &  0.734  &  0.787  \\
22	  &    1.448  &    1.242    &   0.810   &  1.037  &  1.082  &  1.176  \\
24	  &    1.162  &    0.733    &   0.754   &  0.845  &  0.829  &  0.794  \\
26	  &    1.037  &    0.833    &   0.568   &  0.717  &  0.792  &  0.822  \\
28	  &    1.071  &    0.526    &   0.541   &  0.537  &  0.445  &  0.610  \\
30	  &    0.709  &    0.855    &   0.657   &  0.588  &  0.794  &  0.758  \\
\cline{1-7}
\end{tabular}
\end{table*}
%\end{landscape}

Finally, we want to emphasize that the rigid rotor approximation that we considered for the N$_2$ molecule is a valid approximation, and the results that we obtained are not expected to alter much by relaxing this approximation. For example, \cite{10.1063/1.475458} computed the collisional cross-sections for N+N$_2$ using both the rigid rotor and vibrating rotor models, and they discovered that there was no significant difference between the cross-sections computed using these models. Consequently, they utilized the rotational excitation cross-sections to assess the inelastic portion of the Boltzmann kernel. Given the similarity between our O+N$_2$ system and Balakrishnan et al.’s  N+N$_2$ system, we can conclude that the inclusion of vibrational excitations would only have a minimal impact on the cross-sections. The strongly forward-peaked DCSs that we observed in our results also serves as an evidence that inelastic collisions that result in vibrational excitations require significant energy transfer and can only occur at large scattering angles. This is a typical outcome that occurs during collisions between atoms and molecules that take place at high collision energies, which are studied in the context of fast atom thermalization \citep{kharchenko1997kinetics, johnson1982atomic, 10.1093/mnras/stz3366, https://doi.org/10.1029/2000JA000085}.

% This is because these conditions result in the attenuation of superthermal oxygen atoms in N$_2$ collisions.
% in CO$_2$-rich planetary atmospheres (and other similar non-thermal environments that are exposed to radiation)
%----------------------------------------------------------------------------------
\section{SUMMARY and DISCUSSION}
In conclusion, we simulated the quantum collision dynamics between an N$_2$ molecule and an oxygen atom at collision energies reaching up to 4 eV in the center-of-mass frame.  We considered three isotopes of oxygen, namely, $^{16}$O($^3$P), $^{17}$O($^3$P), and $^{18}$O($^3$P). To describe the electronic interactions between N$_2$ and O($^3$P), we utilized the electronic potential energy surfaces that correspond to the first asymptote of the O-N$_2$ complex interactions, as outlined in \cite{koner2020accurate}'s research. To model the N$_2$ molecule, we employed the rigid rotor approximation. Accordingly, the collisions were treated as non-reactive. To simplify the computational complexity, we utilized the coupled state (CS) approximation, as proposed by \cite{doi:10.1063/1.1681388}. Thus, we constructed the cross-sections and related quantities entirely from first principles, without the use of any fitting parameters. The \textit{ab initio} approach, despite being computationally demanding, offers several notable benefits over classical theory. It allows for the assessment of all state-to-state transitions caused by collisions and provides a more comprehensive handling of purely quantum-mechanical effects, which enhances the accuracy and reliability of the models. 

We computed velocity-dependent state-to-state and overall integral cross-sections for elastic and inelastic collisions. We also computed the corresponding differential cross-sections, for all three oxygen isotopes. The elastic cross-sections that we obtained are lower than the elastic cross-sections reported by \cite{https://doi.org/10.1029/98JA02198}. For superthermal energies (> 1 eV), the elastic cross-sections are within the range of 1.96 $\times$ 10$^{-14}$ cm$^2$ at 1 eV collision energy and decreased to 1.37 $\times$ 10$^{-14}$ cm$^2$ at 4 eV collision energy. The computed DCSs indicate a significant preference for forward scattering, with the degree of anisotropy becoming more pronounced as the collision energy increases. %The N$_2$ molecule experiences mainly elastic scattering with minimal energy transfer to its internal degrees of freedom during small-angle forward scattering. On the other hand, if the scattering angle is larger, the total DCSs are almost 100 times higher than the elastic DCSs. This suggests that the transfer of kinetic energy to internal excitations is highly efficient. At the greatest collision energies examined, the amount of total cross-section is roughly half that of the elastic cross-section. This indicates that upto 50\% of the translational kinetic energy of the O($^3$P) atom is transformed into internal excitations, particularly rotations, of the N$_2$ molecule. 
Please note that one should consider a Maxwell-Boltzmann distribution of rotational states of N$_2$ corresponding to the background temperature of the gas through which the superthermal O is being transported. Indeed, we calculated the S-matrices for all state-to-state pairs, and we can provide the state-to-state cross-sections for any combination of initial and final states upon reasonable request.
Further, we find that the difference between the elastic scattering for 0 $\rightarrow$ 0 and e.g., 2 $\rightarrow$ 2 (in $j$ $\rightarrow$ $j'$ notation) is not that large, but it is noticeable, as expected \citep{https://doi.org/10.1029/2012GL050904}.
Overall, the use of the O + N$_2$ cross-sections presented in this study would greatly improve the estimation of photochemical hot O escape rates from planets like Mars and Venus.

\section*{Acknowledgements}
 Sanchit thanks CQuICC IIT Madras for Postdoctoral fellowship. Sumit acknowledges IIT Madras for research fellowship through HTRA.  N. E. K  and M.G. thank the ASPIRE Award for Research Excellence (AARE20-000329-00001) for their funding. N. E. K is partially supported by the Space and Planetary Science Center at Khalifa University through the award 8474000336‐KU‐SPSC. M.G. is partially suppored by Khalifa University award 8474000362. S. S. R. K. C. Y. acknowledges the financial support from IIT Madras through the MPHASIS faculty fellowship and through its new faculty support grants NFSG (IP2021$/$0972CY$/$NFSC008973), NFIG (RF2021$/$0577CY$/$NFIG008973), and DST-SERB (SRG$/$2021$/$001455).

%%%%%%%%%%%%%%%%%%%%%%%%%%%%%%%%%%%%%%%%%%%%%%%%%%
\section*{Data Availability}
The data underlying this article will be shared upon reasonable request to the corresponding author.

\section*{Author contributions}
Sanchit Kumar: Conceptualization, Methodology, Software, Validation, Formal analysis, Investigation, Data curation, Writing – original draft, Visualization; Sumit Kumar: Data curation, Investigation; Marko Gacesa: Conceptualization, Validation; Nayla El-Kork: Conceptualization; Sharma S.R.K.C. Yamijala: Writing – review \& editing, Resources, Supervision, Project administration, Funding acquisition.


\bibliographystyle{mnras}
\bibliography{1} % if your bibtex file is called example.bib

%\begin{thebibliography}{99}
%\end{thebibliography}

%%%%%%%%%%%%%%%%%%%%%%%%%%%%%%%%%%%%%%%%%%%%%%%%%%

%%%%%%%%%%%%%%%%% APPENDICES %%%%%%%%%%%%%%%%%%%%%

%\appendix


%it can be placed in an Appendix which appears after the list of references.. 

%\begin{table} %[h]
%\caption{\label{tab:table3}Elastic cross-sections ($\nu$=0, $j$=0) (units of 10$^{-16}$ cm$^2$) as a function of selected energies (in eV) for $^{17}$O with N$_2$}.
%\centering
%\begin{tabular}{ccccc}
%\cline{1-5}
%\\
 % E (eV)       &   $^3$A' &  1 $^3$A'' & 2 $^3$A" &  Average  \\ \cline{1-5} \\

 %    \cline{1-5}          
%\end{tabular}
%\end{table}

%\begin{table} %[h]\caption{\label{tab:table3}Elastic cross-sections ($\nu$=0, $j$=0) (units of 10$^{-16}$ cm$^2$) as a function of selected energies (in eV) for $^{18}$O with N$_2$}.\centering\begin{tabular}{ccccc}\cline{1-5}\\ E (eV)       &   $^3$A' &  1 $^3$A'' & 2 $^3$A" &  Average  \\ \cline{1-5} \\    \cline{1-5}          \end{tabular}\end{table}
%%%%%%%%%%%%%%%%%%%%%%%%%%%%%%%%%%%%%%%%%%%%%%%%%%

% Don't change these lines
\bsp	% typesetting comment
\label{lastpage}
\end{document}